\documentclass[english,groupedaddress, superscriptaddress,aps,pre,10pt, nofootinbib, notitlepage,longbibliography]{revtex4-2}
\usepackage{mathtools}
\usepackage{amsmath}
\usepackage{amssymb}
\usepackage{bigints}
\usepackage{graphicx}
\usepackage{gensymb}
\usepackage{float}
\usepackage[normalem]{ulem}
\usepackage{comment}
\makeatletter

\renewcommand*{\p@subsection}{}

\renewcommand*{\p@subsubsection}{}  
\makeatother
\usepackage{geometry}
\usepackage{bigints}
\usepackage{graphicx}
\usepackage{gensymb}
\usepackage{mathtools}
\usepackage{amsmath}
\usepackage{amssymb}
\usepackage[T1]{fontenc}
\usepackage[utf8]{inputenc}

\DeclarePairedDelimiter\abs{\lvert}{\rvert}%
\DeclarePairedDelimiter\norm{\lVert}{\rVert}%

\usepackage{color}
\definecolor{blue}{rgb}{0,0,1}
\definecolor{black}{rgb}{0,0,0}

\definecolor{dgreen}{rgb}{0,0.5,0}

\definecolor{dred}{rgb}{0.5,0,0}

\definecolor{dyellow}{rgb}{0.75,0.75,0}

\makeatletter
\let\oldabs\abs
\def\abs{\@ifstar{\oldabs}{\oldabs*}}
\let\oldnorm\norm
\def\norm{\@ifstar{\oldnorm}{\oldnorm*}}
\makeatother

\makeatletter
\usepackage{enumerate}%
\usepackage[caption=false]{subfig}

\begin{document}

\title{Spatial analysis of tails of air pollution PDFs in Europe}

\author{Hankun He}
\affiliation{School of Mathematical Sciences, Queen Mary University of London, London E1 4NS, United Kingdom}

\author{Benjamin Schäfer}
\affiliation{Institute for Automation and Applied Informatics, Karlsruhe Institute of Technology, 76344 Eggenstein-Leopoldshafen, Germany}

\author{Christian Beck}
\affiliation{School of Mathematical Sciences, Queen Mary University of London, London E1 4NS, United Kingdom}

\begin{abstract}
Outdoor air pollution is estimated to cause a huge number of premature deaths worldwide, it catalyses many diseases on a variety of time scales, and it has a detrimental effect on the environment. In light of these impacts it is necessary to obtain a better understanding of the dynamics and statistics of measured air pollution concentrations, including temporal fluctuations of observed concentrations and spatial heterogeneities. Here we present an extensive analysis for measured data from Europe. The observed probability density functions (PDFs) of air pollution concentrations depend very much on the spatial location and on the pollutant substance. 
We analyse a large number of time series data from 3544 different European monitoring sites and show that the PDFs of nitric oxide ($NO$), nitrogen dioxide ($NO_2$) and particulate matter ($PM_{10}$ and $PM_{2.5}$) concentrations generically exhibit heavy tails. These are asymptotically well approximated by $q$-exponential distributions with a given entropic index $q$ and width parameter $\lambda$. We observe that the power-law parameter $q$ and the width parameter $\lambda$ vary widely for the different spatial locations. 
We present the results of our data analysis in the form of a map that shows which parameters $q$ and $\lambda$ are most relevant in a given region. A variety of interesting spatial patterns is observed that correlate to properties of the geographical region. We also present results on typical time scales associated with the dynamical behaviour.
\end{abstract}

\maketitle
\makeatother

\section{Introduction}
Outdoor air pollution is estimated to cause 4.2 million premature deaths per year worldwide, according to estimates of the WHO, and is thus a major killer. It also causes many diseases on a variety of time scales~\cite{WHOPollution,Shah}. There are highly detrimental effects for the environment, as air pollution affects vegetation, natural ecosystems, and plays a role in climate change~\cite{Ortiz}. In all densely populated regions of the word, it represents a major environmental health risk~\cite{health2019state,world2018global,gbd2017global}, impacting human health, ecosystems, climate, infrastructure, and the economy~\cite{Ortiz}. There are many different types of air pollutants and they can interact in a complex way. Two air pollutants, namely particulate matter ($PM$) and nitrogen oxides ($NO_x$), pose a considerable threat to the health of citizens. In 2018 about $55000$ and $417000$ premature deaths in 41 European countries were attributed to $NO_2$ and $PM_{2.5}$, respectively~\cite{Ortiz}. In this contribution to Climate Informatics 2024 we focus on $NO_x$ and $PM$, but our statistical data analytics methods can be applied to other substances as well.

The impact of air pollution on health does not only depend on the pollutant type but also on the type of surrounding area. The EU~\cite{european2011directive} uses environmental area types to classify air quality monitoring sites into traffic, industrial, background, urban, suburban and rural, based on predominant emission sources and building density. From an air quality policy perspective, this allows for evaluating the effectiveness of measures targeting specific emissions sectors, assessing the impact of those pollutants which dominate the area surrounding a given monitoring station.

A lot of air pollution research concentrates onto mean values, but having a thorough understanding of the time-varying statistics, i.e. of the entire probability density function (PDF) of air pollution concentrations is crucial for policymakers involved in defining thresholds or reducing overall exposure to air pollution by suitable mitigation measures. It is also crucial for the construction of suitable statistical physics-based models, using stochastic methods. PDFs such as gamma, log-normal and Weibull distributions~\cite{Hsin} have been widely used for fitting air pollutant concentration data. However, these distributions decay approximately like exponential functions at large values, while previous investigations have found heavy tails in air pollution statistics \cite{williams_superstatistical_2020,code}, which are not well-described by the above distributions. 

Previous papers~\cite{aswin2023lockdown,BALDASANO2020140353,environmental2020effect} have also explored the COVID-19 lockdown effects on air quality (for example in megacities such as London and Delhi), focusing on comparing the PDFs or given moments of the PDFs before and during the lockdown. Superstatistical methods, originating from turbulence modelling~\cite{PhysRevLett.98.064502} and applied to many fields \cite{metzler2020superstatistics, beck2020nonextensive, ourabah2024superstatistics}, offer a powerful effective approach to describe the dynamics of air pollution. These types of models are based on the assumption of well-separated time scales~\cite{williams_superstatistical_2020}. Air pollutants such as $NO_x$ have been described successfully using a superstatistical approach, taking into account nonequilibrium situations with fluctuating variance parameters \cite{williams_superstatistical_2020}. However, the approach in that paper was verified for limited data sets only, chosen from the UK (London), and also only for a limited set of pollutants, namely $NO$ and $NO_2$ only. 
For European data, recent progress was made in~\cite{code},
where it was shown that many measured air pollution time series data can be
generically understood and modelled by superstatistical methods.
In this conference proceedings, we follow up from that approach. We consider a very large set of measured time series data in Europe (3544 locations), extract the best fitting parameters at each spatial location, and then plot a map of Europe where the best fitting parameters are visualized at each point. The results are quite useful to know as they provide some information on what type of power law exponent is to be expected at a given location, thus describing the risk of high-pollution situations occurring at that location.

This paper is organized as follows. We first describe the data set considered and briefly comment on our method of finding the best fitting parameters. Some relevant background on superstatitical methods and $q$-exponential functions will be given. We then present our main result on air pollution statistics in Europe, in form of the visual display of the map of Europe with the given power-law statistics parameters that we obtained in our detailed analysis. Finally, we discuss the observed patterns and spatial correlations.

\section{Methods}
\subsection{The data set considered}
We use air quality monitoring data from a large number of locations in Europe through the interface "Saqgetr", which is an R package available on the Comprehensive R Archive Network (CRAN)~\cite{saqgetr}. Most of the data are openly available from the European Commission’s Airbase and air quality e-reporting (AQER) repositories~\cite{EU_Parliament_Council,EU_Commission}. We also import surface meteorological data from NOAA ISD~\cite{smith2011integrated} via the "worldmet"~\cite{carslaw2017worldmet} R package. Initially, we started with 9698 different locations and used measured data from January 2017 to December 2021, recorded at 1-hour intervals. However, with certain selection criteria applied on the quality of the measured data, this reduces to a set containing 3544 sites. Our selection criteria are described in detail in \cite{code}. A main criterium was that at least one year of measured data should be available.
Our interest is in the behaviour of the tails of the PDFs, as illustrated in Fig.~1. These tails correspond to high pollution states and are most damaging to health. The measuring stations are divided into three categories: traffic, industrial, and background based on predominant emission sources; the surrounding areas are classified as urban, suburban, or rural based on the building density. Measuring station types are combined with area types to provide an overall station classification, and we analyse our data conditioned on this station classification. 

\begin{figure}[t]
\centering
\includegraphics[width=0.7\textwidth]{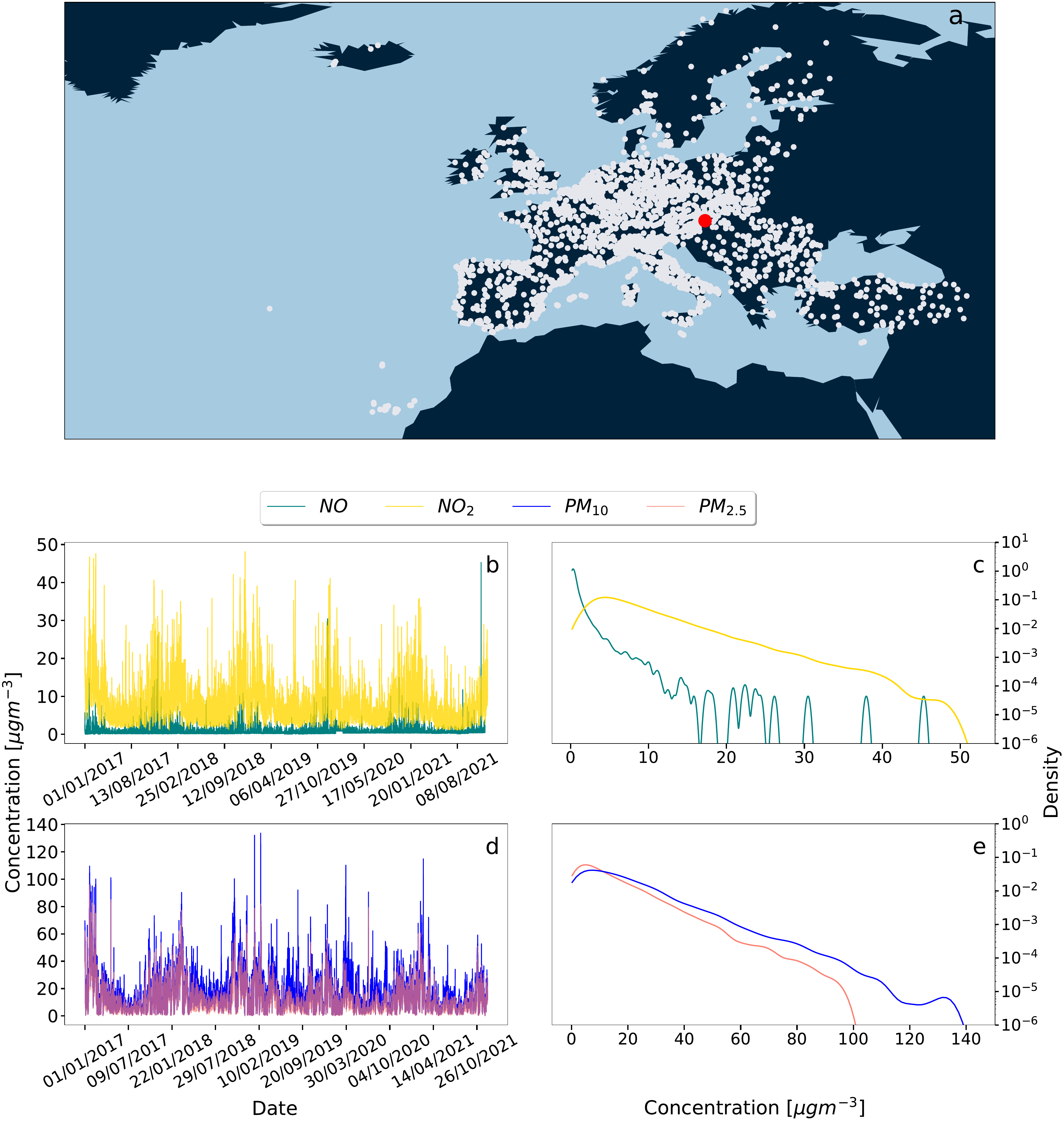}
\caption{(a) The white dots show the positions of all European measuring stations considered. A red dot singles out an example of a station at Illmitz, Austria. For such a given example location, the time series data of the four measured concentrations are shown in panels b and d. The corresponding tails of the PDFs (as shown in c, e) are then automatically analysed with a fitting program \cite{code} that extracts the parameters $(q, \lambda)$ for the best-fitting $q$-exponential. This is done for all 3544 sites}
\end{figure}

\subsection{Fitting with $q$-exponentials}

In \cite{code} evidence was presented that generic air pollution PDFs often decay as a power law. To extract the power law exponent, it is best to apply a fitting approach based on $q$-exponential functions, since these asymptotically approach a power law (with exponent
$-1/(q-1)$ if $q>1$) but also take into account next-to leading order terms in the asymptotics. The form of $q$-exponentials is motivated by generalized versions of statistical mechanics based on non-additive entropies
\cite{tsallis2009introduction}, as well as on superstatistical approaches \cite{beck-cohen, BCS}. The reason for the occurrence of $q$-exponentials is that exponentials with varying rate parameters, when integrated over the probability density of the rate parameter, in leading order lead to $q$-exponentials. The functional form of the $q$-exponential is given by
\begin{equation}
f_{q,\lambda}(x) = (2 - q) \lambda [1 -\lambda (1 - q) x]^\frac{1}{1 - q} \text{ for } 1 -\lambda (1 - q) x \geq 0, x>0, 
\label{eq:qexp}
\end{equation}
where $q$ is the entropic index~\cite{tsallis2009introduction,hanel2011,jizba2019}, $\lambda$ is a positive width parameter and $x$, in our case, denotes the air pollutant concentration. Eq.~\eqref{eq:qexp} contains the exponential distribution as a special case, namely that is obtained in the limit
$q \to 1$,
as the $q$-exponential function, defined as $e_{q}(x)= [1 + (1 - q) x]^\frac{1}{1 - q}$, converges to the exponential function in the limit $q\rightarrow 1$. For $q<1$, $f_{q,\lambda}(x)$  is also well defined but in this case lives on a finite support and becomes exactly zero above a critical value x, since, by definition, $e_{q}(x) = 0$ for $1-\lambda(1-q)x<0$. In contrast, if $q > 1$, $1 -\lambda(1-q)x>0$, then Eq.~\eqref{eq:qexp} exhibits power-law asymptotic behavior. Both cases are relevant for air pollution fits. 

The occurrence of $q$-exponentials with $q>1$ in PDFs of
complex systems has been previously explained as originating from a superstatistical dynamics \cite{beck-cohen, BCS, ourabah2024superstatistics}.
In these types of models, one assumes a temporally fluctuating parameter $\hat{\lambda}$ for local exponential distributions $\sim e^{-\hat{\lambda}x}$. This means that in a sense the parameters of the system are random variables as well, but they change on a much larger time scale. These variations of $\hat{\lambda}$ take place
on a long time scale which is much longer than the local relaxation time to equilibrium. Physically, one may think of this as describing changing weather conditions (e.g. changes in the wind pattern) at the measuring site.
The marginal distribution, obtained by integration over all possible values of $\hat{\lambda}$, and
describing the long-term behaviour of the air pollution concentration dynamics, is then a  $q$-exponential, with 
\begin{equation}
    q= \frac{ \langle \hat{\lambda}^2 \rangle}{\langle \hat{\lambda} \rangle^2}.
    \label{eq:q}
\end{equation}
Here $\langle \cdots \rangle$ denotes the expectation
with respect to the PDF of $\hat{\lambda}$, see \cite{beck-cohen} for more details. Strictly speaking, a $q$-exponential is only obtained exactly if $\hat{\lambda}$ is $\Gamma$-distributed,
but the general idea of superstatistics is
that a parameter $q$ can be defined by Eq.~\eqref{eq:q} for more general distributions different than the $\Gamma$ distribution as well.

\begin{figure}[h]
    \centering
    \includegraphics[width=0.6\textwidth]{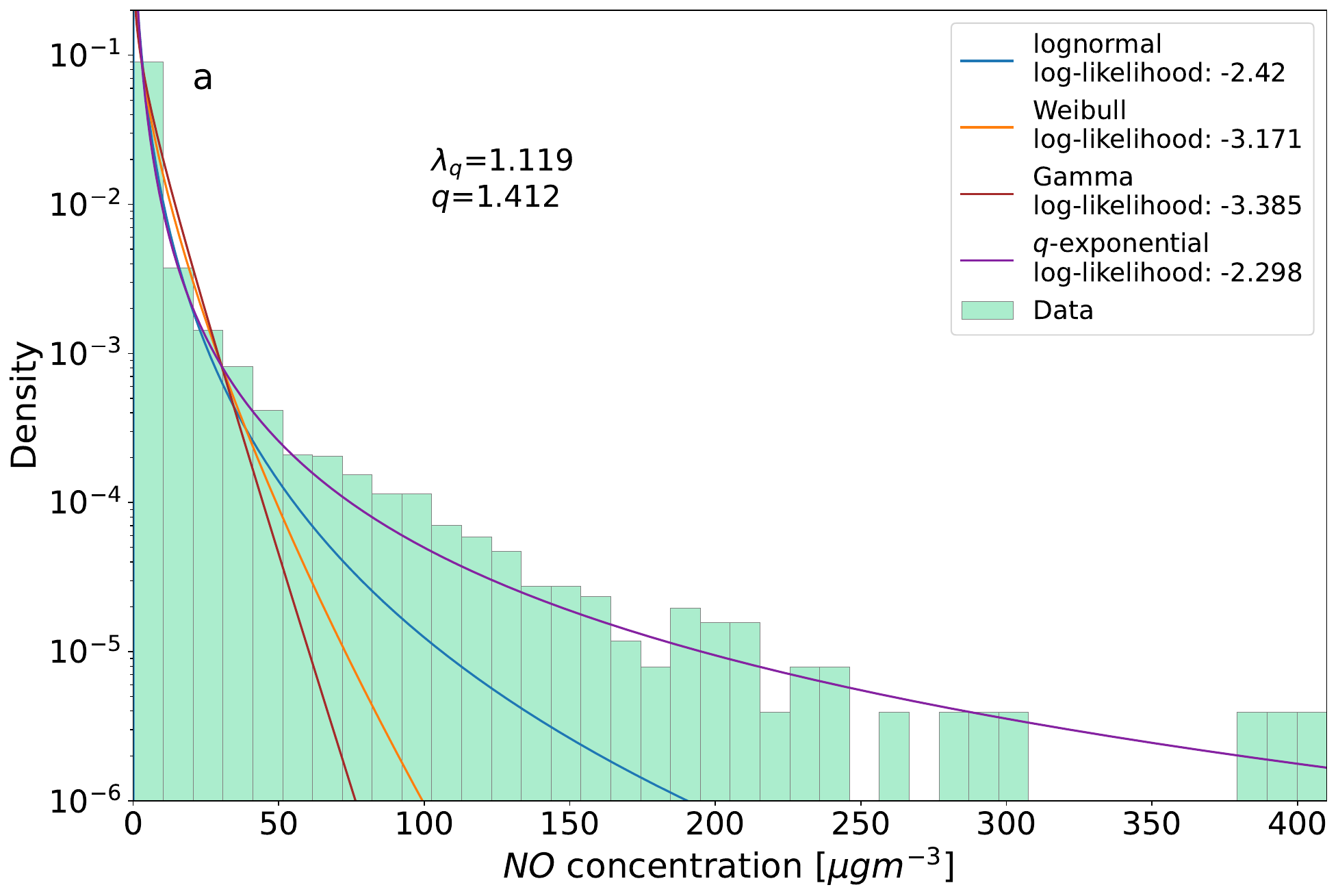}
    \caption{Measured histogram of $NO$ concentrations, recorded at the example site Barnsley Gawber, UK, together with the best log-normal (blue), Weibull (orange), Gamma (brown) and $q$-exponential (purple) fits, together
    with their respective log-likelihood values. Also displayed are the optimum values of the $q$ and $\lambda$ parameters for the $q$-exponential distribution. The $q$-exponential fits the measured data best, as indicated by the highest log-likelihood value. Similar plots can be produced for all 3544 sites}
    \label{fig:fittings}
\end{figure}

In our automatically applied fitting procedure for the 3544 measuring stations, we obtain the optimal
fitting parameters via MLE, i.e. maximizing the likelihood under the assumption that the data originates
from the assumed $q$-exponential distribution.
One may ask whether other distributions, such as the previously studied lognormal distribution, Weibull distribution, or gamma distribution might yield equally good fits as the $q$-exponential distribution. To illustrate the goodness of fit, we compare all these functions with the measured histogram for $NO$ data in Fig.~2 for an example site (Barnsley Gawber in the UK). Clearly, the $q$-exponential fit yields the best fit of the tails of the distribution. It also yields the largest log-likelihood value compared to the other distributions. We systematically compared the goodness-of-fit using maximum likelihood estimation to determine the best fitting parameters and hence computed the likelihood for each fitting function. Overall, we find that $q$-exponentials generate the highest log-likelihoods for all four pollutants, when averaging over all sites. Hence, we apply $q$-exponentials as our main method for analyzing the tails of air pollution statistics.
See also github~\cite{code} for full technical details.

\subsection{Extracting the long superstatistical time scale $T$}

Utilizing the $q$-exponential distribution for fitting the PDFs opens the option to identify a {\em long time scale} $T$, as discussed in the superstatistics approach \cite{beck-cohen,BCS}. In this view, we assume that on time scales shorter than $T$, we observe random samples from a simple distribution, which we assume to be exponential distributions here. Meanwhile, the overall complex time series, characterized frequently by heavy-tailed probability distributions, is seen as a superposition of several of these simpler, local distributions.


When considering shorter time slices, if the local time series approximately follows an exponential distribution, the kurtosis for a local snapshot should be $K_{\text{exp}} = 9$. To determine $T$, we compute the local average kurtosis \cite{BCS} as follows:

\begin{equation}
\overline{\kappa} (\Delta t)= \frac{1}{t_{max}-\Delta t} \int_{0}^{t_{max}-\Delta t} d t_0 \frac{{\langle (u-\overline{u})^{4} \rangle}_{t_0,\Delta t}}{{\langle (u-\overline{u})^{2} \rangle^2}_{t_0,\Delta t}} ,
\end{equation}
where $t_{max}$ is the length of the time series, ${\langle \rangle}_{t_0,\Delta t}$ deotes the expectation for the time slice of length $\Delta t$ starting at $t_0$. The long time scale $T$ is then defined by the condition $\overline{\kappa} (T) = K_{exp}$. In this case,  the average kurtosis of windows of length $T$ has a kurtosis $K_{exp}=9$. Once $T$ is determined, the time series can be divided into multiple snapshots, each spanning a length of $T$. This provides a collection of slices with approximately local exponential distributions, each with a distinct decay parameter $\hat{\lambda}$ that fluctuates on the time scale $T$. 

\label{chi_sq}

\section{Results}

Fig.~3 shows the best-fitting parameters $q$ and $\lambda$ for the tails of the observed PDFs at all 3544 sites evaluated. As mentioned above,
it is natural to expect $q$-exponential distributions as
these simply arise from the agglomeration of many exponential distributions that have temporal fluctuations of the effective decay rate.

\begin{figure}[t]
\centering
    \includegraphics[width=0.8\textwidth]{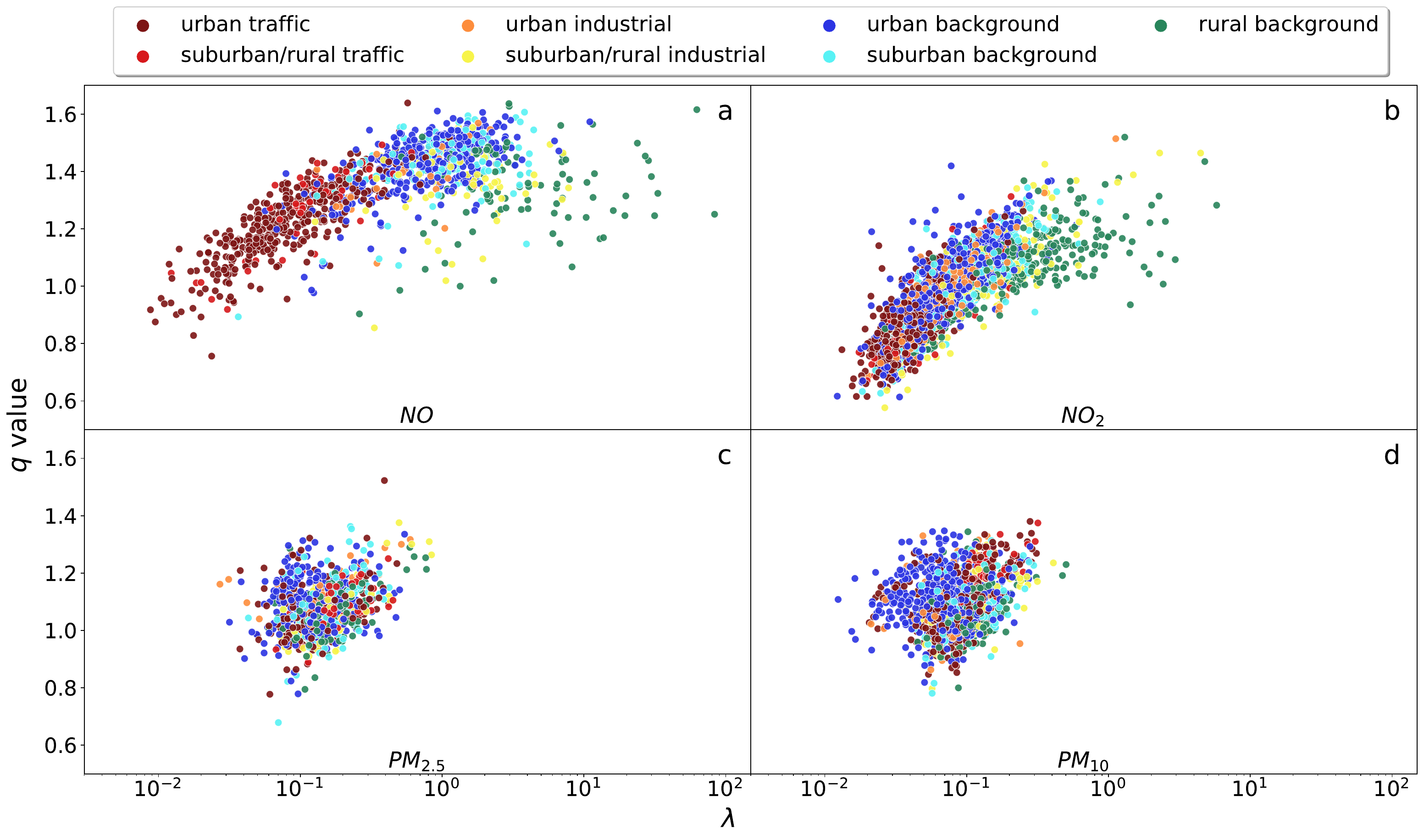}
    \caption{Best-fitting parameters of $q$-exponentials at the various measuring stations. There is an increasing trend of $q$ versus $\log \lambda$ for $NO$ (a) and $NO_2$ (b), whereas a disk-shaped pattern is observed for $PM_{2.5}$ (c) and $PM_{10}$ (d). The colors encode the area type where the measurements were done. There are predominant patches of a single color in the parameter space for $NO$ and $NO_2$, where for $PM_{2.5}$ and $PM_{10}$ the color pattern looks more mixed. Deep green points correspond to cleaner rural areas.} 
    \label{fig:q_vs_l}
\end{figure}

What is noticeable is the fact that one observes an immensely large range of values of the parameter $\lambda$ for the best-fitting $q$-exponential as given in Eq.~(\ref{eq:qexp}) for the various measuring stations. Note the logarithmic scale of the plots. The parameter $\lambda$ can take on values as small as $10^{-2}$ up to values as large as $10^2$, which spans four orders of magnitude. Typical $q$-values are in the range $0.8-1.4$, but there are subtle differences between the various substances, with $NO$ reaching large $q$-values such as 1.6, and $NO_2$ reaching small $q$-values such as $0.6$ in the scattering plots. Figure \ref{fig:q_vs_l}, panels a and b, indicate that roughly $q$ grows linearly with $\log \lambda$ for $NO_x$. Traffic areas are clustered on the left, while urban/suburban background points are in the middle part, and rural background points are scattered widely at the right. The PDF decay rate increases as $\lambda$ increases from highly polluted urban traffic sites to less polluted rural background areas. For $PM_{2.5}$ and $PM_{10}$, there is a slightly different statistical relationship between $q$ values and $\lambda$, as can be seen in Fig.\ref{fig:q_vs_l} panels c and d. The urban points approach small $\lambda$ values such as $0.01$ for $PM_{10}$, and suburban/rural traffic, suburban/rural industrial and rural background points cluster at the right hand side with large $\lambda$. The observed range of $\lambda$ values is smaller as compared to the case of $NO_x$, and the shape of possible values $(q, \lambda)$ as displayed in the Figure is more spherical. The stronger colour mixing for $PM$ may be due to the fact that by air movement there is transport of the pollutants, hence the distributions cannot be uniquely identified with the original environmental types where the $PM$-particles were produced. 

So far we conditioned the observed distributions on the environmental type, such as high traffic region, industrial region, or rural area. This leads to a pretty mixed pattern of states as displayed in Fig.~4. However, it is useful to keep the spatial information where the actual measurements were done. Thus, the idea in the following is to plot an air pollution map of Europe where the
color encodes the value of the best-fitting fitting parameters $(q, \lambda)$ relevant in that given region of Europe. This map is shown in Fig.~5. The novel approach here is that we do not only consider average values of pollution concentrations as observed in a given region but encode information on the entire probability distribution. The parameter $q$ describes the power-law tails of the probability densities, which decay with an exponent given by $-1/(q-1)$. Thus $q$ is related to the frequency of occurrence of extreme events describing high-pollution states.
\begin{figure}[t]
\centering
\includegraphics[width=0.75\textwidth]{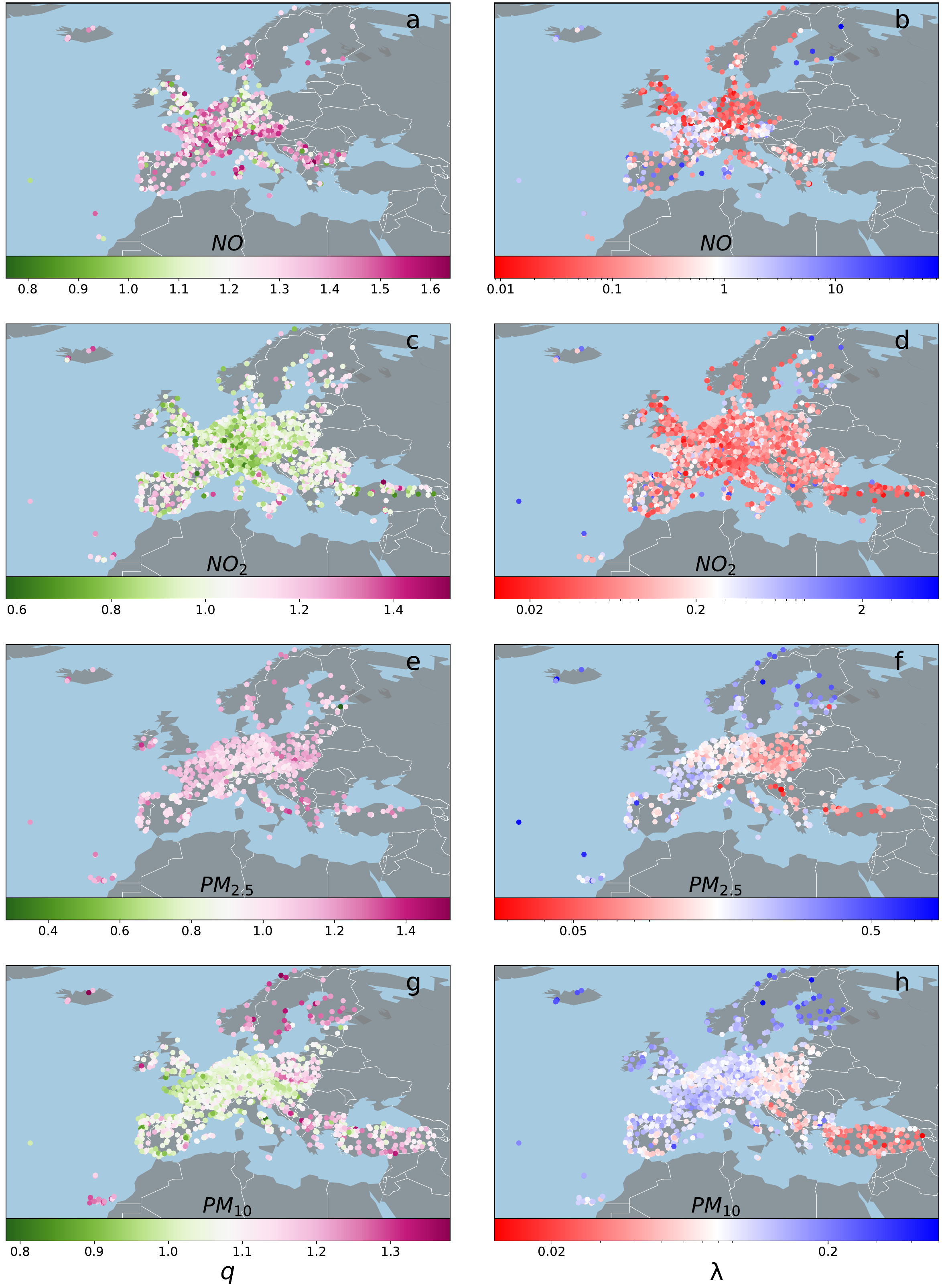}
\caption{Spatial distribution of best-fitting parameters $(q,\lambda)$ characterizing the measured PDFs of $NO_x$ and $PM_x$ pollutants across Europe. The color codes are directly indicated in the individual figures. A large value of $q$ indicates heavy tails in the distribution. A small value of $\lambda$ indicates heavy average pollution. The pollutant characteristics across Europe is quite inhomogeneous}
\label{q_l_map}
\end{figure}

The $q$ values for $NO_2$ are smaller in Germany and the UK as compared to other countries, in fact below 1 (green colors) which indicates there are no heavy tails. Since the primary emission source of $NO_x$ is fuel combustion, this pattern suggests that these regions cope better with extreme situations and have a decreased susceptibility to extreme $NO_x$ events. The $q$ values for $PM$ pollutants exhibit a slightly different spatial pattern, with lower values (green) in the Western regions and higher values (red) in the East. 
The $\lambda$ values in Central Europe and the UK are smaller compared to those in other European regions, indicating a higher average pollution,
in particular for $NO$.
As for the spatial distribution of $\lambda$ values for $PM$ pollutants, there is a slight eastward shift in the clustering of red dots, suggesting a higher pollution in Eastern Europe. 

Note that the average concentration of pollutants is proportional to $\lambda^{-1}$. This means the smaller $\lambda$ the bigger is the average air pollution concentration in that region. For this reason, we have color-coded smaller values of $\lambda$ in red and bigger values in blue. Some of the Scandinavian countries appear to host a couple of blue dots, whereas in Turkey there are big concentrations of red dots, in particular for $PM_{10}$. For the $q$-values, describing extreme events, the tail in the distribution is more pronounced the bigger $q$ is. Some Eastern European countries exhibit particularly heavy tails, as indicated by red colors in the $q$ panels. On the other hand, if $q$ is smaller than 1 then the distributions live on a finite support and there are no heavy tails.
This case is indicated by green colors in the $q$-plane. Some regions of Germany and Western Europe exhibit predominantly green colors.

\begin{figure}[t]
\centering
\includegraphics[width=0.75\textwidth]{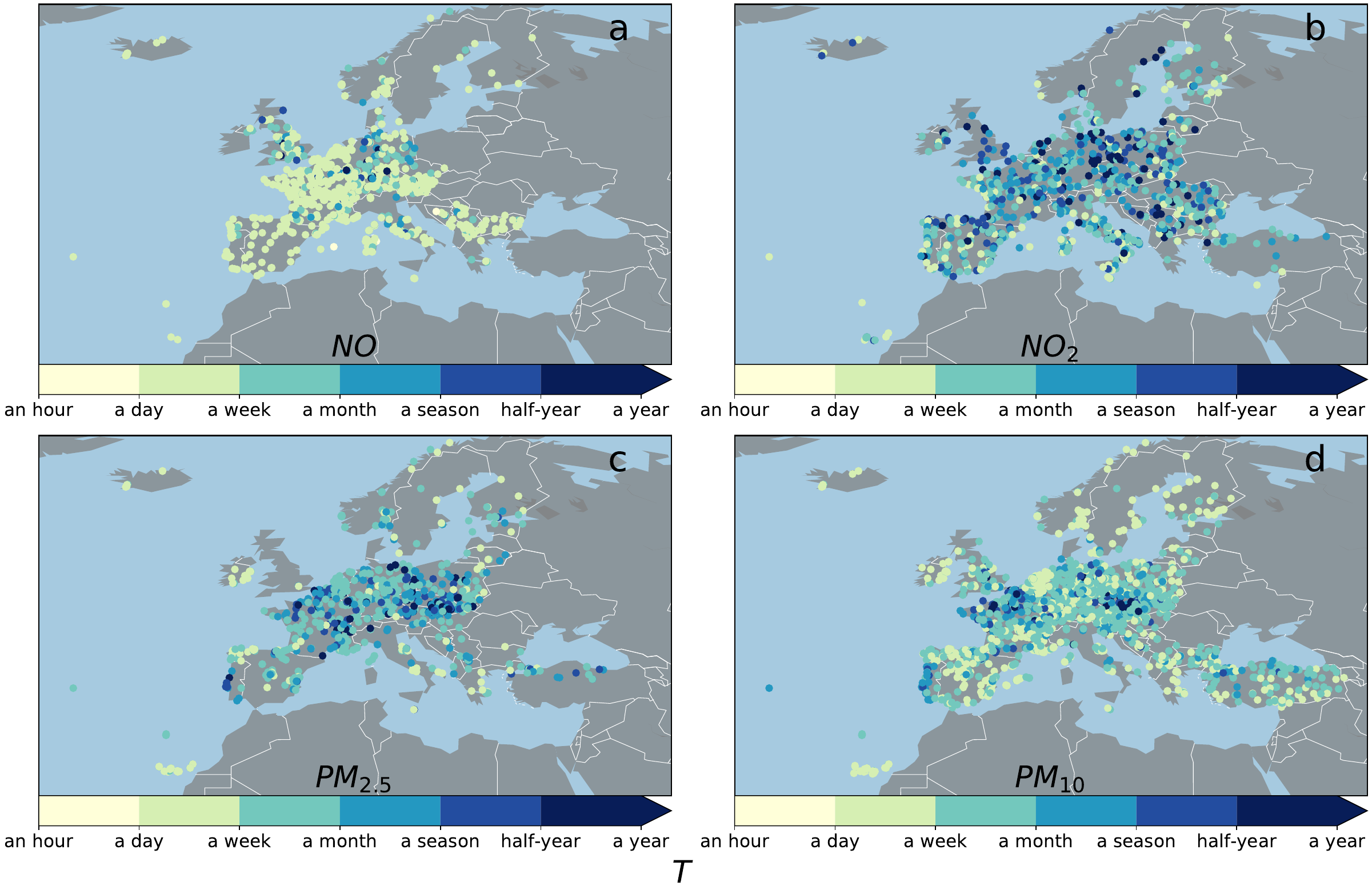}
\caption{The long time scales $T$ describing the scale of typical changes of the temporal mean and variance of the measured time series for $NO$ (a), $NO_2$ (b), $PM_{2.5}$ (c) and $PM_{10}$ (d). 
The color coding is explained at the bottom of each figure}
\label{long_scale_map}
\end{figure}

Fig. \ref{long_scale_map} displays the superstatistical time scale $T$ as extracted for the four pollutants, indicating distinct temporal dynamics. $NO$ overall has a majority of shorter time scales $T$, typically less than a week. $NO_2$, on the other hand, comprises higher values of time scales $T$, mostly ranging between a week to a year. This difference could be explained by the fact that $NO$ is highly reactive and tends to quickly oxidize to form $NO_2$ in the presence of oxygen. This rapid transformation means that $NO$ concentrations can change swiftly over short periods. 
The extracted time scales for $PM_{2.5}$ are of similar order of magnitude as for $NO_2$, whereas the extracted time scales for the $PM_{10}$ dynamics appear to be smaller than for $PM_{2.5}$ on average.

\section{Discussion}
In this paper, we applied data analytics tools to better understand the probability densities of measured air pollution time series.
In particular, we were interested in the tail behaviour. The parameters
$q$ and $\lambda$, as obtained from the optimum fittings of data from 3544 measuring stations in Europe, exhibit interesting patterns in the $(q, \lambda)$ plane. They also exhibit interesting patterns when associated with a map of Europe. There are significant differences between the $NO$, $NO_2$ and the $PM$ statistics. 

The shape of scattered $(q, \lambda)$ values in the plane depends on the local characteristics, e.g. whether the measurements are done in a rural area or densely populated area. There is a typical range of parameter values for a given class of this type. We observe an immensely large range of values of the parameter $\lambda$ for the various measuring stations. 

We also observe a significant spatial heterogeneity in the best-fitting shape parameters of the PDFs.
Eastern Europe shows a tendency for more severe $PM$ pollution events and higher average concentration exposure, described by higher $q$ values and lower $\lambda$ values, respectively. Northern Europe shows a tendency for lower air pollutant concentration, described by higher $\lambda$ values, but is more susceptible to extreme events, described by higher $q$ values. Additionally, there are strong variations of
PDFs at local level as well. These disparities highlight the importance of tailored strategies for air pollution mitigation strategies in different areas. Recall that $q$ contains information about the tail, i.e. extreme events and $\lambda$ about the scale and thereby the mean pollution level, i.e. we distinguish thereby between regions with low and high average pollution and simultaneously regions with many or few extreme events.

We also extracted the long superstatistical time scales $T$ for the measured time series for each of the 3544 measuring sites and for each of the four pollutants considered, thus incorporating dynamical information that amends the static information contained in the PDFs. This information was displayed in form of a map of Europe, too, and shows a variety of time scales relevant at different spatial locations. The observed wide range of time scales, in particular for $NO_2$ and $PM_{2.5}$, can be attributed to a variety of  factors, including but not limited to fluctuating local meteorological conditions and anthropogenic activities.

Future research, for each given location taking into account the local circumstances, should help to gain a clearer systematic understanding of the relationship between meteorological and anthropogenic factors and the relevant time scales and $(q,\lambda)$-values of air pollutants. A main result of our investigation is strong local heterogeneity, i.e. the temporal dynamics of air pollution can be quite different if one proceeds from one local measurement station to the next. Different local wind conditions are expected to play an important role in this context.

Further statistical analysis of more detailed data at local level could support policymakers to produce more precise rules and thresholds for individual types of environmental conditions and meteorological conditions, taking into account fluctuations, extreme events, and variations of concentrations on different time scales. 

For our statistical analysis the different types of air pollutants have different data availability in the various geographical regions. For this reason, in the air pollution maps that we have produced in this paper, some regions are less frequently covered by points than others. We hope that in the future more measurements at further locations will become available, given the importance of the air pollution problem. Our analysis could also be extended to other substances, such as sulfur oxide, carbon oxide and ozone, besides $NO_x$ and $PM$. We stress again that the detailed description of the entire pollution PDF, including the exact behaviour of the tails, seems critical here to better estimate the risks of very high pollution situations.

\subsection*{Data Availability Statement}
Data sets analyzed in the present study can be obtained via the saqgetr package from \url{https://github.com/skgrange/saqgetr}. The code to generate the figures in the paper, as well as the implementation of the method for the data sets used in the paper, are available at \url{https://github.com/hurst0415/Spatial-heterogeneity-of-air-pollution-statistics}. 

\bibliographystyle{naturemag}
\bibliography{main}

\begin{thebibliography}{10}
\expandafter\ifx\csname url\endcsname\relax
  \def\url#1{\texttt{#1}}\fi
\expandafter\ifx\csname urlprefix\endcsname\relax\def\urlprefix{URL }\fi
\providecommand{\bibinfo}[2]{#2}
\providecommand{\eprint}[2][]{\url{#2}}

\bibitem{WHOPollution}
\bibinfo{author}{{World Health Organization (WHO)}}.
\newblock \bibinfo{title}{How air pollution is destroying our health} (\bibinfo{year}{2020}).
\newblock \urlprefix\url{https://www.who.int/news-room/spotlight/how-air-pollution-is-destroying-our-health}.

\bibitem{Shah}
\bibinfo{author}{Shah, A.~S.} \emph{et~al.}
\newblock \bibinfo{title}{Short term exposure to air pollution and stroke: systematic review and meta-analysis}.
\newblock \emph{\bibinfo{journal}{BMJ}} \textbf{\bibinfo{volume}{350}}, \bibinfo{pages}{h1295} (\bibinfo{year}{2015}).

\bibitem{Ortiz}
\bibinfo{author}{Ortiz, A.}, \bibinfo{author}{Guerreiro, C.}, \bibinfo{author}{Soares, J.} \& \bibinfo{author}{Agency, E.~E.}
\newblock \emph{\bibinfo{title}{Air Quality in Europe: 2020 Report}}.
\newblock EEA report (\bibinfo{publisher}{Publications Office of the European Union}, \bibinfo{year}{2020}).
\newblock \urlprefix\url{https://books.google.co.uk/books?id=swtezgEACAAJ}.

\bibitem{health2019state}
\bibinfo{author}{{Health Effects Institute}}.
\newblock \bibinfo{title}{State of global air 2019} (\bibinfo{year}{2019}).

\bibitem{world2018global}
\bibinfo{author}{{World Health Organization}} \emph{et~al.}
\newblock \bibinfo{title}{Who global ambient air quality database (update 2018)}.
\newblock \emph{\bibinfo{journal}{World Health Organization: Geneva, Switzerland}}  (\bibinfo{year}{2018}).

\bibitem{gbd2017global}
\bibinfo{author}{{GBD 2016 Risk Factors Collaborators}} \emph{et~al.}
\newblock \bibinfo{title}{Global, regional, and national comparative risk assessment of 84 behavioural, environmental and occupational, and metabolic risks or clusters of risks, 1990--2016: a systematic analysis for the global burden of disease study 2016}.
\newblock \emph{\bibinfo{journal}{Lancet (London, England)}} \textbf{\bibinfo{volume}{390}}, \bibinfo{pages}{1345} (\bibinfo{year}{2017}).

\bibitem{european2011directive}
\bibinfo{author}{{European Parliament}}.
\newblock \bibinfo{title}{{Directive 2011/24/EU of the European Parliament and of the Council of 9 March 2011 on the application of patients’ rights in cross-border healthcare}} (\bibinfo{year}{2011}).

\bibitem{Hsin}
\bibinfo{author}{Lu, H.-C.}
\newblock \bibinfo{title}{Comparisons of statistical characteristic of air pollutants in {Taiwan} by frequency distribution, journal of the air \& waste management}.
\newblock \emph{\bibinfo{journal}{Association}} \textbf{\bibinfo{volume}{53:5}}, \bibinfo{pages}{608--616} (\bibinfo{year}{2003}).

\bibitem{williams_superstatistical_2020}
\bibinfo{author}{Williams, G.}, \bibinfo{author}{Schäfer, B.} \& \bibinfo{author}{Beck, C.}
\newblock \bibinfo{title}{Superstatistical approach to air pollution statistics}.
\newblock \emph{\bibinfo{journal}{Physical Review Research}} \textbf{\bibinfo{volume}{2}}, \bibinfo{pages}{013019} (\bibinfo{year}{2020}).
\newblock \urlprefix\url{https://link.aps.org/doi/10.1103/PhysRevResearch.2.013019}.
\newblock \bibinfo{note}{Publisher: American Physical Society}.

\bibitem{code}
\bibinfo{author}{He, H.}
\newblock \bibinfo{title}{Spatial heterogeneity of air pollution statistics}.
\newblock \bibinfo{howpublished}{\url{https://github.com/hurst0415/Spatial-heterogeneity-of-air-pollution-statistics}} (\bibinfo{year}{2024}).
\newblock \bibinfo{note}{Available online}.

\bibitem{aswin2023lockdown}
\bibinfo{author}{Aswin~Giri, J.} \emph{et~al.}
\newblock \bibinfo{title}{Lockdown effects on air quality in megacities during the first and second waves of covid-19 pandemic}.
\newblock \emph{\bibinfo{journal}{Journal of The Institution of Engineers (India): Series A}} \textbf{\bibinfo{volume}{104}}, \bibinfo{pages}{155--165} (\bibinfo{year}{2023}).

\bibitem{BALDASANO2020140353}
\bibinfo{author}{Baldasano, J.~M.}
\newblock \bibinfo{title}{{COVID-19 lockdown effects on air quality by NO2 in the cities of Barcelona and Madrid (Spain)}}.
\newblock \emph{\bibinfo{journal}{Science of The Total Environment}} \textbf{\bibinfo{volume}{741}}, \bibinfo{pages}{140353} (\bibinfo{year}{2020}).
\newblock \urlprefix\url{https://www.sciencedirect.com/science/article/pii/S0048969720338754}.

\bibitem{environmental2020effect}
\bibinfo{author}{{Environmental Research Group}} \emph{et~al.}
\newblock \bibinfo{title}{{The effect of COVID-19 lockdown measures on air quality in London in 2020}} (\bibinfo{year}{2020}).

\bibitem{PhysRevLett.98.064502}
\bibinfo{author}{Beck, C.}
\newblock \bibinfo{title}{{Statistics of Three-Dimensional Lagrangian Turbulence}}.
\newblock \emph{\bibinfo{journal}{Phys. Rev. Lett.}} \textbf{\bibinfo{volume}{98}}, \bibinfo{pages}{064502} (\bibinfo{year}{2007}).
\newblock \urlprefix\url{https://link.aps.org/doi/10.1103/PhysRevLett.98.064502}.

\bibitem{metzler2020superstatistics}
\bibinfo{author}{Metzler, R.}
\newblock \bibinfo{title}{{Superstatistics and non-Gaussian diffusion}}.
\newblock \emph{\bibinfo{journal}{The European Physical Journal Special Topics}} \textbf{\bibinfo{volume}{229}}, \bibinfo{pages}{711--728} (\bibinfo{year}{2020}).

\bibitem{beck2020nonextensive}
\bibinfo{author}{Beck, C.} \emph{et~al.}
\newblock \bibinfo{title}{Nonextensive statistical mechanics, superstatistics and beyond: theory and applications in astrophysical and other complex systems} (\bibinfo{year}{2020}).

\bibitem{ourabah2024superstatistics}
\bibinfo{author}{Ourabah, K.}
\newblock \bibinfo{title}{Superstatistics from a dynamical perspective: Entropy and relaxation}.
\newblock \emph{\bibinfo{journal}{Physical Review E}} \textbf{\bibinfo{volume}{109}}, \bibinfo{pages}{014127} (\bibinfo{year}{2024}).

\bibitem{saqgetr}
\bibinfo{author}{Grange, S.~K.}
\newblock \bibinfo{title}{saqgetr r package} (\bibinfo{year}{2019}).
\newblock \urlprefix\url{https://github.com/skgrange/saqgetr}.

\bibitem{EU_Parliament_Council}
\bibinfo{author}{{European Parliament, Council of the European Union}}.
\newblock \bibinfo{title}{{DIRECTIVE 2008/50/EC OF THE EUROPEAN PARLIAMENT AND OF THE COUNCIL of 21 May 2008 on ambient air quality and cleaner air for Europe}} (\bibinfo{year}{2008}).
\newblock \urlprefix\url{https://eur-lex.europa.eu/legal-content/EN/ALL/?uri=CELEX%3A32008L0050}.

\bibitem{EU_Commission}
\bibinfo{author}{{European Commission}}.
\newblock \bibinfo{title}{{2011/850/EU: Commission Implementing Decision of 12 December 2011 laying down rules for Directives 2004/107/EC and 2008/50/EC of the European Parliament and of the Council as regards the reciprocal exchange of information and reporting on ambient air quality (notified under document C(2011) 9068)}} (\bibinfo{year}{2011}).
\newblock \urlprefix\url{https://eur-lex.europa.eu/legal-content/EN/TXT/?uri=CELEX%3A32011D0850}.

\bibitem{smith2011integrated}
\bibinfo{author}{Smith, A.}, \bibinfo{author}{Lott, N.} \& \bibinfo{author}{Vose, R.}
\newblock \bibinfo{title}{The integrated surface database: Recent developments and partnerships}.
\newblock \emph{\bibinfo{journal}{Bulletin of the American Meteorological Society}} \textbf{\bibinfo{volume}{92}}, \bibinfo{pages}{704--708} (\bibinfo{year}{2011}).

\bibitem{carslaw2017worldmet}
\bibinfo{author}{Carslaw, D.}
\newblock \bibinfo{title}{{Worldmet: import surface meteorological data from NOAA integrated surface database (ISD)}}.
\newblock \emph{\bibinfo{journal}{GitHub repository}}  (\bibinfo{year}{2017}).
\newblock \urlprefix\url{https://github.com/davidcarslaw/worldmet}.

\bibitem{tsallis2009introduction}
\bibinfo{author}{Tsallis, C.}
\newblock \bibinfo{title}{Introduction to nonextensive statistical mechanics: approaching a complex world}.
\newblock \emph{\bibinfo{journal}{Springer}} \textbf{\bibinfo{volume}{1}}, \bibinfo{pages}{2--1} (\bibinfo{year}{2009}).

\bibitem{beck-cohen}
\bibinfo{author}{Beck, C.} \& \bibinfo{author}{Cohen, E. G.~D.}
\newblock \bibinfo{title}{Superstatistics}.
\newblock \emph{\bibinfo{journal}{Physica A: Statistical mechanics and its applications}} \textbf{\bibinfo{volume}{322}}, \bibinfo{pages}{267--275} (\bibinfo{year}{2003}).

\bibitem{BCS}
\bibinfo{author}{Beck, C.}, \bibinfo{author}{Cohen, E. G.~D.} \& \bibinfo{author}{Swinney, H.~L.}
\newblock \bibinfo{title}{From time series to superstatistics}.
\newblock \emph{\bibinfo{journal}{Physical Review E}} \textbf{\bibinfo{volume}{72}}, \bibinfo{pages}{056133} (\bibinfo{year}{2005}).

\bibitem{hanel2011}
\bibinfo{author}{Hanel, R.}, \bibinfo{author}{Thurner, S.} \& \bibinfo{author}{Gell-Mann, M.}
\newblock \bibinfo{title}{Generalized entropies and the transformation group of superstatistics}.
\newblock \emph{\bibinfo{journal}{Proceedings of the National Academy of Sciences}} \textbf{\bibinfo{volume}{108}}, \bibinfo{pages}{6390--6394} (\bibinfo{year}{2011}).

\bibitem{jizba2019}
\bibinfo{author}{Jizba, P.} \& \bibinfo{author}{Korbel, J.}
\newblock \bibinfo{title}{Maximum entropy principle in statistical inference: Case for non-shannonian entropies}.
\newblock \emph{\bibinfo{journal}{Physical Review Letters}} \textbf{\bibinfo{volume}{122}}, \bibinfo{pages}{120601} (\bibinfo{year}{2019}).

\end{thebibliography}

\subsection*{Author Contributions}
Conceptualization - All authors; Methodology - All authors; Data curation - H.H.; B.S. Data visualisation - All authors; Writing original draft - All authors. All authors approved the final submitted draft.

\subsection*{Competing Interests}
The authors declare no competing interests.

\subsection*{Funding Statement}
We gratefully acknowledge funding from a 2024 QMUL ISPF grant awarded to C.B. and funding from the Helmholtz Association and the Networking Fund through Helmholtz AI and under grant no. VH-NG-1727 to B.S.

\subsection*{Provenance Statement}
This article is part of the Climate Informatics 2024 proceedings and was accepted in Environmental Data Science on the basis of the Climate Informatics peer review process.

\end{document}